\documentclass[12pt]{article}
\usepackage{latexsym}
\usepackage{amsfonts}
\usepackage{epsfig}

\catcode`\@=11

\def\section{\@startsection{section}{1}{\z@}{3.5ex plus 1ex minus
 .2ex}{2.3ex plus .2ex}{\bf}}

\def\thesubsection{\arabic{section}.\arabic{subsection}}
\renewcommand{\subsection}[1]{\addtocounter{subsection}{1}
\vspace{2.5mm}\par\noindent {\it \thesubsection . #1}\par
 \vspace{0.5mm} }
\catcode`\@=12 \mathchardef\varGamma="0100
\mathchardef\varDelta="0101 \mathchardef\varTheta="0102
\mathchardef\varLambda="0103 \mathchardef\varXi="0104
\mathchardef\varPi="0105 \mathchardef\varSigma="0106
\mathchardef\varUpsilon="0107 \mathchardef\varPhi="0108
\mathchardef\varPsi="0109 \mathchardef\varOmega="010A
\def\bfone{\relax{\rm 1\kern-.35em 1}}
\DeclareFontFamily{U}{rsf}{} \DeclareFontShape{U}{rsf}{m}{n}{
  <5> <6> rsfs5 <7> <8> <9> rsfs7 <10-> rsfs10}{}
\DeclareMathAlphabet\Scr{U}{rsf}{m}{n}
\begin{document}
\begin{titlepage}
\thispagestyle{empty}
\begin{flushright}
\hfill{CERN-PH-TH/2004-245}\\
\hfill{DESY-04-234}
\end{flushright}
\vspace{35pt}
\begin{center}{ \LARGE{\bf
Scalar potential for the gauged Heisenberg algebra and a non--polynomial antisymmetric tensor theory}}\\
 \vspace{50pt} {\bf R. D'Auria$^\S$, S.
Ferrara$^\sharp$, M. Trigiante$^\S$, S. Vaul\`a$^\natural$}\\
\vspace{15pt}
$\S${\it Dipartimento di Fisica, Politecnico di Torino \\
C.so Duca degli Abruzzi, 24, I-10129 Torino\\
Istituto Nazionale di Fisica Nucleare, Sezione di Torino,
Italy}\\[1mm] {E-mail: riccardo.dauria@polito.it,
mario.trigiante@polito.it}\\
\vspace{15pt} $\sharp${\it CERN, Physics Department, CH 1211
Geneva 23, Switzerland.}\\ {\it INFN, Laboratori Nucleari di
Frascati, Italy}\\[1mm] {E-mail:
sergio.ferrara@cern.ch}\\
 \vspace{15pt}
$\natural${\it DESY, Theory Group\\
Notkestrasse 85, Lab. 2a D-22603 Hamburg, Germany\\
II. Institut f\"ur Theoretische Physik,\\
Luruper Chaussee 149, D-22761 Hamburg, Germany}\\[1mm] {E-mail:
silvia.vaula@desy.de} \vspace{15pt}
\begin{abstract}
We study some issues related to the effective theory of Calabi--Yau compactifications 
with fluxes in Type II theories. At first the scalar potential for a generic electric abelian gauging of the Heisenberg algebra, underlying all possible gaugings of RR isometries, is presented and shown to exhibit, in some circumstances, a ``dual'' no--scale structure under the interchange of hypermultiplets and vector multiplets. Subsequently a new setting of such theories, when all RR scalars are dualized into antisymmetric tensors, is discussed. This formulation falls in the class of non--polynomial tensor theories considered long ago by Freedman and Townsend and it may be relevant for the introduction of both electric and magnetic charges.
\end{abstract}
\end{center}
\end{titlepage}
\newpage
\baselineskip 6 mm
\section{Introduction}

In a recent paper \cite{dftv} we have proposed a representation of Calabi--Yau
compactifications with fluxes as gauged supergravity where the
gauge group is embedded in the Heisenberg algebra of the $2\,
h_1+3$ isometries \cite{fs,dwvp1,dwvp2} (recall that according to our conventions
$h_1=h_{11},\,h_2=h_{12}$ in Type IIB setting while
$h_1=h_{12},\,h_2=h_{11}$ in Type IIA) corresponding to the R--R
scalars and the scalar field $a$ dual to the NS 2--form
$B_{\mu\nu}$. The Heisenberg algebra has the following symplectic
structure:
\begin{eqnarray}
[X_A,\,X_B]&=&-2 \, {\mathbb C}_{AB}\,{\Scr Z}\,,\label{heis}
\end{eqnarray}
where $X_A=\{X^\Lambda,\,X_\Lambda\}$, $\Lambda=1,\dots, h_1+1$
and ${\mathbb C}_{AB}$ is the symplectic invariant matrix:
${\mathbb C}^\Lambda{}_\Sigma=-{\mathbb
C}_\Sigma{}^\Lambda=\delta^\Lambda{}_\Sigma$,
 ${\mathbb C}_{\Lambda\Sigma}={\mathbb C}^{\Lambda\Sigma}=0$. For the sake of
 simplicity it will be useful in the sequel to extend the index
 $A$ to an index $m=\{0,A\}=0,\dots, h_1+1$, so that, defining $X_0={\Scr
 Z}$, equation (\ref{heis}) can be rewritten in the form
 $[X_m,\,X_n]=f_{mn}{}^p\,X_p$, where the only non vanishing
 components of the structure constants are:
 $f_{AB}{}^0=-2\,{\mathbb C}_{AB}$.
\par
It is straightforward to see that gauging this algebra by $h_2$
vectors corresponds to defining the following gauge generators:
\begin{eqnarray}
T_I&=&e_I{}^A\, X_A+c_I\, {\Scr Z}=e_I{}^m\, X_m\,,
\end{eqnarray}
where $e_I{}^0=c_I$ and $I=0,\dots , h_2$. The gauge algebra
$\{T_I\}$ closes within the Heisenberg algebra, with structure
constants $f_{IJ}{}^K$, namely
 $[T_I,\,T_J]=f_{IJ}{}^K\,T_K$, only if the following conditions
 are satisfied:
 \begin{eqnarray}
e_I{}^m\, e_J{}^n\,f_{mn}{}^p&=&f_{IJ}{}^K\,e_K{}^p\,,\nonumber \\
f_{IJ}{}^K\,e_K{}^A&=&0\,.
 \end{eqnarray}
In what follows we shall need to restrict ourselves to an abelian
gauge algebra, namely $[T_I,\,T_J]=0$. In this case the ``electric'' charges $e_I{}^m$ satisfy the trivial--cocycle condition \cite{dftv}:
 \begin{eqnarray}
e_I{}^m\, e_J{}^n\,f_{mn}{}^k&=&-2\, e_I{}^A\, e_J{}_A=0\,.
 \end{eqnarray}
The $N=2$ scalar potential for such general electric gauging is then constructed using
the standard formulae of $N=2$ supergravity \cite{magri}.   
On the other hand if the scalar $a$ is dualized into the 2--form
$B_{\mu\nu}$, the residual algebra associated to the generators
$X_A$ becomes abelian: $[X_A,\,X_B]=0$. This reflects the fact
that in the Lagrangian all the scalar fields
$V_A=\{\zeta^\Lambda,\,\tilde{\zeta}_\Lambda\}$ parametrizing
$X_A$ appear covered by derivatives and the coupling to the $B$
field
\begin{eqnarray}
dV_A\wedge dV^A\wedge B\,,\label{vvb}
\end{eqnarray}
is invariant under $V_A\rightarrow V_A+{\rm const.}$ and
$B\rightarrow B+d\Lambda$.\par When the Lagrangian is written in
such form one can use Poincar\'e duality between 1 and 3-forms in
four dimensions to obtain a dual Lagrangian which involves tensor
fields $B_m=\{B^\Lambda,\,B_\Lambda, B_0\}$, where $B_0=B$,
$B_\Lambda$ are dual to $\zeta^\Lambda$  and $B^\Lambda$ dual to $\tilde{\zeta}_\Lambda$. 
The coupling (\ref{vvb}) gives
rise to a non--polynomial theory when Poincar\'e duality is used
and this reflects the non--abelian structure of the original
Heisenberg algebra written in (\ref{heis}). This makes contact
with an old work by Freedman and Townsend \cite{ft} where a
non--abelian tensor gauge theory was formulated and also the
coupling to gauge fields (associated to a semisimple gauge group) discussed.\par 
The paper is organized as
follows: \par In section 2 we exhibit the form of the scalar potential
for a generic electric (abelian) gauging of the Heisenberg
algebra and obtain a structure which, for some
particular choice for the electric fluxes, reproduces the scalar
potential derived in the context of Type IIA,B compactifications \cite{bcf,m,bghl,cklt,da,lm,kk,dkpz}
on \emph{half--flat} manifolds with NS fluxes \cite{glmw,gm,halfflat}. \par
 A dual structure is found in the
potential when the Heisenberg algebra of the special quaternionic
isometries is gauged. This allows to prove, as an example, that in
the case in which $e_I{}^\Lambda=0$ the potential for $e_{I
\Lambda}=e_{I 0}$, with cubic special geometry for the vector
sector, has the same no--scale structure as the potential for $e_{I
\Lambda}=e_{0 \Lambda}$ and cubic special geometry for the
hypermultiplets. This allows to prove the equivalence between
compactifications on a half--flat Calabi--Yau manifold and
compactifications in the presence of NS fluxes on the mirror
manifold. The antisymmetric tensor  formulation should allow to describe
theories were both electric and magnetic charges are introduced \cite{ddsv,df,ls},
thus realizing  a completely ${\rm Sp}(2\,h_2+2)\times {\rm
Sp}(2\,h_1+2)$ duality invariant scalar potential, which contain
all the previous examples as particular cases.
\section{Scalar potential with electric fluxes}
Let us start by introducing some notations. The scalars of a
special quaternionic manifold are denoted by:
\begin{eqnarray}
q^u&=&\{\phi,\,a,\,\zeta^\Lambda,\,\tilde{\zeta}_\Lambda,\,Z^\Lambda\}\,\,\,;\,\,\,\,\Lambda=0,\dots,
h_1\,,
\end{eqnarray}
where, from Type IIB point of view, $a$ is the scalar dual to the
2--form NS tensor $B_{\mu\nu}$, $\zeta^0=C_{(0)}$,
$\zeta^\Lambda=C_{(2)}^\Lambda$ ($\Lambda\neq 0$), $\tilde{\zeta}_0$ is dual to
$C_{\mu\nu}$, $\tilde{\zeta}_\Lambda=C_{(4)\,\Lambda}$ ($\Lambda\neq 0$), $\phi$ is
the four--dimensional dilaton and $Z^\Lambda$ is the projective special coordinate vector
describing the remaining NS moduli $z^a$ ($a=1,\dots, h_1$):
$Z^0=1,\,Z^a=z^a$. As anticipated in the introduction, there exists a subgroup of the isometry group
generated by a Heisenberg algebra $\{X^\Lambda,\,X_\Lambda,{\Scr
Z} \}$, whose action of the hyperscalars has the following form:
\begin{eqnarray}
\delta\zeta^\Lambda&=&\alpha^\Lambda\,,\nonumber\\
\delta\tilde{\zeta}_\Lambda&=& \beta_\Lambda\,,\nonumber\\
\delta a &=&
\gamma+\alpha^\Lambda\tilde{\zeta}_\Lambda-\beta_\Lambda\zeta^\Lambda\,,
\label{transf}\end{eqnarray} and whose structure is\footnote{Note that, when extending the Heisenberg algebra with 
the Cartan generator corresponding to the $\phi$ coordinate, one obtains the $2\, h_1+4$ dimensional solvable algebra generating the symmetric coset ${\rm SU}(1,2+h_1)/{\rm U}(1)\times {\rm SU}(2+h_1)$ \cite{adfft}. This is the minimal number of isometries of a special quaternionic manifold.}:
\begin{eqnarray}\label{Heis}
\left[X^\Lambda,\, X_\Sigma\right]&=&-2\,\delta^\Lambda{}_\Sigma\,
{\Scr
Z}\,,\nonumber\\
\left[X^\Lambda,\, X^\Sigma\right]&=&\left[X_\Lambda,\,
X_\Sigma\right]=\left[X^\Lambda,\,{\Scr
Z}\right]=\left[X_\Lambda,\, {\Scr Z}\right]=0\,.
\end{eqnarray}
 The gauge
generators $T_I$ ($I=0,\dots, h_2$) have the following form in
terms of the Heisenberg isometries:
\begin{eqnarray}\label{embe1}
T_I&=&e_I{}^\Lambda\, X_\Lambda-e_{I\Lambda}\, X^\Lambda+c_I\,
{\Scr Z}\,,
\end{eqnarray}
and the corresponding Killing vectors are:
\begin{eqnarray}
k_I&=&(c_I+e_I{}^\Lambda\,\tilde{\zeta}_\Lambda-e_{I\Lambda}\,\zeta^\Lambda)\,\frac{\partial}{\partial
a }+e_I{}^\Lambda\,\frac{\partial}{\partial \zeta^\Lambda
}+e_{I\Lambda}\,\frac{\partial}{\partial \tilde{\zeta}_\Lambda
}\,.
\end{eqnarray}
The general form of the ${\Scr N}=2$ scalar potential is \cite{magri}:
\begin{equation}
{\Scr V} = 4\,  h_{uv} k^u_I k^v_J \,L^I\,\overline{L}^J+
g_{i\bar{\jmath}}\, k^i_I k^{\bar{\jmath}}_J \,L^I\,
\overline{L}^J+ (U^{IJ}-3\,L^I \overline{L}^J){\Scr P}^x_I\,{\Scr
P}^x_J\,,\label{pot}
\end{equation}
where $L^I$ are the upper components of the covariantly holomorphic symplectic section of the special K\"ahler manifold, 
$k^u_I$ and $k^i_I$ are respectively the quaternionic and special K\"ahler Killing vectors and $U^{IJ}$ denotes the matrix \cite{magri}:
\begin{eqnarray}
U^{IJ}&=&-\frac{1}{2}\,{\rm Im}({\Scr N})^{-1\, IJ}-\overline{L}^I\,L^J=D_iL^J\, D_{\bar{j}}\overline{L}^J\,g^{i\bar{j}}\,,
\end{eqnarray}
$h_{uv},\,g_{i\bar{j}}$ being the metrics of the hyper-- and vector multiplet geometries.
The second term in (\ref{pot}) does not contribute to the gauging we are
considering, which involves quaternionic  isometries only. The
expression for the momentum maps ${\Scr P}^x_J$ is \cite{g,dff}:
\begin{eqnarray}
{\Scr P}^x_I&=&k_I^u\,\omega^x_u\,,\label{p}
\end{eqnarray}
where $\omega^x$ is the ${\rm SU}(2)$ connection. This form is
Heisenberg--invariant and so is therefore the ${\rm SU}(2)$
curvature. This justifies the absence of a compensator on the
right hand side of eq. (\ref{p}).\par The NS scalars $Z^\Lambda$
parametrize a special K\"ahler submanifold of the quaternionic
manifold. It is useful to characterize this manifold in terms of a
K\"ahler potential $K$, a prepotential ${\Scr F} $ and period matrix
\footnote{Note that the K\"ahler potential $K_{SK}$ and period matrix ${\Scr N}_{SK}$, defined according to the conventions for special K\"ahler manifolds adopted for instance in \cite{magri}, are related to the corresponding quantities in (\ref{mn}) in the following way: $e^{K_{SK}}=\frac{e^K}{4}\,\,\,\,;\,\,\,{\Scr
N}_{SK}=\frac{1}{4}\,{\mathcal M}$}:
\begin{eqnarray}
{\mathcal
M}&=& i\,\overline{{\Scr N}}_s\,,\label{mn}
\end{eqnarray}
where ${\Scr N}_s$ is the period matrix as defined in \cite{fs}
and ${\mathcal M}$ is the matrix used in e.g. \cite{glmw} and $K$
being the K\"ahler potential associated with it.
 Let us define the following forms:
\begin{eqnarray}
v&=&e^{\tilde{K}}\,[d\phi-i\,(da+\tilde{\zeta}^T\,
d\zeta-\zeta^T\, d\tilde{\zeta})]\,,\nonumber\\
u&=&2i\,e^{\frac{\tilde{K}+\hat{K}}{2}}\,
Z^T\,(\overline{{\mathcal
M}}\,d\zeta+d\tilde{\zeta})\,, \nonumber\\
E&=&i\,e^{\frac{\tilde{K}-\hat{K}}{2}}\,P\,N^{-1}\,(\overline{{\mathcal
M}}\,d\zeta+d\tilde{\zeta})\,, \nonumber\\
e&=&P\,dZ\,,\label{uvforms}
\end{eqnarray}
where
\begin{eqnarray}
e^{\tilde{K}}&=&\frac{1}{2\phi}=\frac{e^{2\,\varphi}}{2},\,\,\,;\,\,\,\,\,e^{\hat{K}}=\frac{1}{2\bar{Z}NZ}=\frac{e^K}{2}\,,
\end{eqnarray}
and $\varphi$ denotes the four dimensional dilaton. The matrices $P$ and $N$ are defined as follows:
\begin{eqnarray}
P^{\underline{\sigma}}{}_0 &=&-e_\lambda{}^{\underline{\sigma}}{} Z^\lambda\,\,\,;\,\,\,\,P^{\underline{\sigma}}{}_\lambda=e_\lambda{}^{\underline{\sigma}}\,\,\,(\lambda=1,\dots, h_1)\,,\\
N_{\Lambda\Sigma}&=&\frac{1}{2}{\rm Re}(\frac{\partial^2 {\Scr F}}{\partial Z^\Lambda\partial Z^\Sigma})\,.
\end{eqnarray}
$e_\lambda{}^{\underline{\sigma}}$ being the vielbein of the special K\"ahler manifold embedded in the quaternionic manifold (the underlined indices being the rigid ones).
\par The metric on
the quaternionic manifold reads \cite{dftv}:
\begin{eqnarray}
ds^2&=& \bar{v}\,v+\bar{u}\,u+\bar{E}\,E+\bar{e}\,e=\nonumber\\&=&
K_{a\bar{b}}\,d
z^a\,d\bar{z}^{\bar{b}}+\frac{1}{4\,\phi^2}\,(d\phi)^2+
\frac{1}{4\,\phi^2}\,(d a+ d V\times V)^2-\frac{1}{2\,\phi}\,d
V\,{\Scr M}\,d V\,,
\end{eqnarray}
where the symplectic matrix ${\Scr M}$ is defined as
follows:\begin{eqnarray} {\Scr M}&=&\left(\matrix{\bfone & R\cr 0
& \bfone }\right)\left(\matrix{I& 0\cr 0 & I^{-1}
}\right)\left(\matrix{\bfone &0\cr R&
\bfone}\right)\,\,\,;\,\,\,\, R={\rm Re} ({\mathcal M}),\,I={\rm
Im} ({\mathcal M})\,,
\end{eqnarray}
and $V$ denotes the symplectic section: $
 V=\left(\matrix{\zeta^\Lambda\cr
\tilde{\zeta}_\Lambda}\right)$.\par It is useful to rewrite the
scalar potential in two equivalent ways:
\begin{eqnarray}
{\Scr V}&=&4\,  h_{uv}\, k^u_I\, k^v_J \,L^I\,\overline{L}^J+
(U^{IJ}-3\,L^I \overline{L}^J)\,k^u_I\,
k^v_J\,\omega^x_u\,\omega^x_v\,,\label{V1}\\
{\Scr V}&=&-\frac{1}{2}\,{\rm Im}{\Scr N}^{-1 IJ}\,k^u_I
k^v_J\,\omega^x_u\,\omega^x_v + 4\,( h_{uv}-
\omega^x_u\,\omega^x_v )\,k^u_I \,k^v_J
\,L^I\,\overline{L}^J\,.\label{V2}
\end{eqnarray}
  In
order to evaluate the expression on the right hand side of eq.
(\ref{V2}) it is useful to compute the following quantity:
\begin{eqnarray}
G_{IJ}&=&k_I^u\,k_J^v\,(h_{uv}-\omega^x_u\,\omega^x_v)=k_I^u\,k_J^v\,
[\bar{v}\,v+\bar{u}\,u+\bar{E}\,E-(\bar{v}\,v+4\,\bar{u}\,u)]_{uv}\,.
\end{eqnarray}
Using the following notation:
\begin{eqnarray}
r_I&=&c_I+2\,(e_{I}{}^{\Lambda}\,\tilde{\zeta}_\Lambda-
e_{I\Lambda}\,\zeta^\Lambda)\,\,\,;\,\,\,\,s_{I\Lambda}=e_{I\Lambda}+e_I{}^\Sigma\,
\overline{{\mathcal M}}_{\Sigma\Lambda}\,,
\end{eqnarray}
we can express $G_{IJ}$ as follows:
\begin{eqnarray}
G_{IJ}&=& 2 \,e^{\tilde{K}}\,\bar{s}_{I\Lambda}\,s_{J\Sigma}\,\left({\Scr
U}-3\,{\mathcal L}^\dagger\,{\mathcal L}
\right)^{\Lambda\Sigma}\,\,;\,\,\,\,{\Scr
U}=-\frac{I}{2}-{\mathcal L}^\dagger\,{\mathcal L}\,\,\,;\,\,\,
{\mathcal L}=e^{\frac{K}{2}}\,Z\,.
\end{eqnarray}
In deriving the above expression for $G_{IJ}$ we made use of the
following properties:
\begin{eqnarray}
N^{-1}P^\dagger P N^{-1}&=&e^{K}\,(-N^{-1}+{\mathcal
L}^T\,\overline{{\mathcal L}})\,,\nonumber\\
-\frac{I}{2}&=&-N^{-1}+{\mathcal L}^T\,\overline{{\mathcal
L}}+{\mathcal L}^\dagger\,{\mathcal L}\,.
\end{eqnarray}
Now we can evaluate the two equivalent expressions for the scalar
potential given in eqs. (\ref{V1}) and (\ref{V2}):
\begin{eqnarray}
{\Scr V}&=&\overline{L}^I\,{L}^J\,\left[\frac{1}{\phi^2}\,(c_I+2\,e_I\times
V)\,(c_J+2\,e_J\times V)-\frac{2}{\phi}\,e_I\,{\Scr
M}\,e_J\right]+\nonumber\\&&+\frac{1}{2\,\phi}\,(U-3\,L^\dagger
{L})^{(IJ)}\,\left(\frac{1}{2\,\phi}\,r_I\,r_J+8\,
\bar{s}_{I\Lambda}\,s_{J\Sigma}\,\overline{{\mathcal L}}^\Lambda
{\mathcal L}^\Sigma \right)\,,\label{V12}\\
{\Scr V}&=&-\frac{1}{4\,\phi}\,{\rm Im}{\Scr N}^{-1
IJ}\,\left(\frac{1}{2\,\phi}\,r_I\,r_J+8\,
\bar{s}_{I\Lambda}\,s_{J\Sigma}\,\overline{{\mathcal L}}^\Lambda
{\mathcal L}^\Sigma \right)+\nonumber\\
&&+\frac{4}{\phi}\,\overline{L}^I\,{L}^J\,\bar{s}_{(I|\Lambda}\,s_{J)\Sigma}\,\left({\Scr
U}-3\,{\mathcal L}^\dagger\,{\mathcal L}
\right)^{\Lambda\Sigma}\,,\label{V22}
\end{eqnarray}
where we have introduced the following vectors:
$e_I=\left(\matrix{e_I{}^\Lambda\cr e_{I\Lambda}}\right)$. The
first equation (\ref{V12}) is useful for those gaugings which
involve just the graviphoton $A^0_\mu$, e.g. Type IIA with NS flux
or Type IIB on a half--flat ``mirror'' manifold. Indeed in these
cases the term in the second line of (\ref{V12}) does not
contribute for cubic special geometries in the vector multiplet
sector since \cite{magri}:
\begin{eqnarray}
(U-3\,L^\dagger {L})^{00}&=&0\,.
\end{eqnarray}
Similarly the expression (\ref{V22}) is of particular use for
those gaugings which involve only isometries $\Lambda=0$, like for
instance Type IIA on a half--flat manifold or Type IIB on the
``mirror'' manifold with NS flux since, for cubic special
quaternionic geometries:
\begin{eqnarray}
\left({\Scr U}-3\,{\mathcal L}^\dagger\,{\mathcal L}
\right)^{00}&=&0\,\,\Rightarrow\,\,\,e^K=-\frac{1}{8}\,
(I^{-1})^{00}.
\end{eqnarray}
In both cases the expressions for the potential reduce to those given in \cite{glmw,gm}.
\section{The non--polynomial antisymmetric tensor Lagrangian}
In this section we discuss the dualization of the special quaternionic $\sigma$--model
when the $2\,h_1+3 $ scalars, parametrizing the Heisenberg algebra are dualized into (2--form) antisymmetric tensors.
Let us start with the Lagrangian for the Heisenberg scalars
($a,\zeta^\Lambda,\,\tilde{\zeta}_\Lambda$) as given in \cite{dftv}.
In order to dualize the scalar $a$ into the NS two form
$B_{\mu\nu}$ we substitute $da$ by the unconstrained 1--form $\eta$ and add a further term:
\begin{eqnarray}
L&=& -\frac{1}{4\,\phi^2}\,(\eta+ d V\times V)\wedge \star (\eta+
d V\times V)+\frac{1}{2\,\phi}\,d V^A\,\wedge \star\,{\Scr
M}_{AB}\, dV^B+\nonumber\\&&H\wedge (\eta-da) \,.
\end{eqnarray}
$H$ being a 3--form Lagrange multiplier.
Integrating out $\eta$ we obtain the dual Lagrangian:
\begin{eqnarray}
L&=& -\phi^2\,H\wedge\star H+\frac{1}{2\,\phi}\,d V^A\,\wedge
\star\,{\Scr M}_{AB}\, dV^B-dV\times dV\wedge B \,.\label{stage1}
\end{eqnarray}
We see that upon dualization of the scalar $a$ all the R--R
scalars appear covered by derivatives, so that we may perform a
further dualization. This is achieved by substituting in
(\ref{stage1}) $dV^A$ by $v^A$, adding the Lagrange multiplier term 
$(v^A-dV^A)\wedge G_{A}$, $G_A$ being a set of 3--forms, and integrating out $v^A$.  We find:
\begin{eqnarray}
\frac{1}{\phi}\,{\Scr M}_{AB}\star v^B+f_{AB}{}^0\, v^B\wedge
B_0+G_A&=&0\,.\label{eqv}
\end{eqnarray}
Denoting by $G_A^\mu=\epsilon^{\mu\nu\rho\sigma}\,
G_{\nu\rho\sigma\,A}$, the above equation in tensor components
reads:
\begin{eqnarray}
G_{\mu A}&=& K_{A\mu,\, B\nu}\,v^{\nu B}\,,\label{gv}\\
 K_{\mu A,\, \nu B}&=&\Delta_{AB}\,
 g_{\mu\nu}-f_{AB}{}^0\,\epsilon_{\mu\nu\rho\sigma}\,
 B_0^{\rho\sigma}\,,\\
 \Delta_{AB}&=&\frac{1}{\phi}\, {\Scr M}_{AB}\,,
\end{eqnarray}
where $B_{\mu\nu\, 0}=B_{\mu\nu}$. Upon implementation of eq.
(\ref{eqv}) the dual Lagrangian can be computed to be:
\begin{eqnarray}
{\Scr L}_D&=&-\phi^2\,H\wedge\star H+\frac{1}{2}\, G_{\mu A}\,
\tilde{K}^{\mu A,\,\nu B}\, G_{\nu B}\,,\nonumber\\
K_{\mu A,\, \nu B}\, \tilde{K}^{B\nu,\, C\rho}&=&\delta^C_A\,
\delta^\rho_\mu\,.\label{ld}
\end{eqnarray}
This Lagrangian has the form of the non--polynomial model discussed in
\cite{ft}. Using the above notations eq. (\ref{gv}) can be
inverted to give:
\begin{eqnarray}
v^{\mu A}&=&\tilde{K}^{A\mu,\, B\nu}\, G_{\nu B}\,.
\end{eqnarray}
We may now derive the equations of motion and the invariance of
the Lagrangian (\ref{ld}). In order to compute the variation of
the Lagrangian corresponding to an arbitrary variation of the
tensor fields $B,\, B_A$, it is convenient to express the various
terms in the variation in terms of the composite fields $v^A$
using equation (\ref{gv}). One then obtains:
\begin{eqnarray}
\delta {\Scr L}_D&=&{\Scr F}^0\wedge \delta B_0+ {\Scr F}^A\wedge
\delta B_A\,,\label{dld}\\
{\Scr F}^0&=& dv^0+\frac{1}{2}\, f_{AB}{}^0\, v^A\wedge v^B=dv^0-
{\mathbb C}_{AB}\, v^A\wedge v^B\,\,;\,\,\,\,{\Scr F}^A=dv^A\,.
\end{eqnarray}
The equations of motion therefore read:
\begin{eqnarray}
{\Scr F}^0&=&{\Scr F}^A=0\,.
\end{eqnarray}
Furthermore, using eq. (\ref{dld}), one can check that the
Lagrangian (\ref{ld}) is invariant, up to total derivatives, under
the following tensor--gauge transformation:
\begin{eqnarray}
\delta B_0&=&D\xi_0=d\xi_0\,\,\,;\,\,\,\,\, \delta
B_A=D\xi_a=d\xi_A+f_{AB}{}^0\, v^B\, \xi_0=d\xi_A- 2\,{\mathbb
C}_{AB}\,\, v^B\, \xi_0\,.\label{db2nd}
\end{eqnarray}
Following \cite{ft}, the dual Lagrangian (\ref{ld}) can be written
in a polynomial form by adopting a first order formalism in which the
fields $v^A$ are treated as independent of $B_A$. In this
formulation the Lagrangian can be written in the following form:
\begin{eqnarray}
{\Scr L}^{(1)}_D&=&-\phi^2\,H\wedge\star H+\frac{1}{2}\,
\Delta_{AB}\,v^A\wedge \star v^B+B_0\wedge {\Scr F}^0+B_A\wedge
{\Scr F}^A\,.
\end{eqnarray}
Indeed the equation of motion for $v^A$ yields the relation
(\ref{gv}) which, if substituted in ${\Scr L}^{(1)}_D$, reproduces
the second order Lagrangian ${\Scr L}_D$. As already noted in
\cite{ft}, the tensor--gauge invariance of the first order
Lagrangian is much simpler since the corresponding variations are
now written in the following form:
\begin{eqnarray}
\delta B_0&=& d\xi_0\,\,\,;\,\,\,\delta B_A=
D\xi_A\,\,\,;\,\,\,\,\delta v^A=0\,,\label{db1st}
\end{eqnarray}
and such gauge transformations are of course abelian.\par
The $N=2$ supersymmetric completion of the non--polynomial Lagrangian (\ref{ld}) is not straightforward and should involve the self--coupling of the $h_1$ double--tensor multiplet with a triple--tensor multiplet (the latter originating from the universal 
hypermultiplet \cite{cfg}).

\subsubsection*{Gauging of the model in the presence of RR fluxes.} We can now try to couple the model to
$h_2+1$ vector fields through electric and magnetic charges
$e_I{}^m=\{e_I{}^0,\,e_I{}^A\}$, $m^{Im}=\{m^{I0},\,m^{IA}\}$. Introducing only  R--R charges $e_I{}^0,\,m^{I0}$ poses no problem
and the gauged Lagrangian has the form:
\begin{eqnarray}
{\Scr L}^{(1)}_g&=&-\phi^2\,H\wedge\star H+\frac{1}{2}\,
\Delta_{AB}\,v^A\wedge \star v^B+B_0\wedge {\Scr F}^0+B_A\wedge
{\Scr F}^A-\nonumber\\
&&-\frac{1}{2}\,{\rm Im}({\Scr N})_{IJ}\,(dA^I+m^{I0}\, B_0)\wedge
\star (dA^I+m^{I0}\, B_0)+\nonumber\\
&&+{\rm Re}({\Scr N})_{IJ}\,(dA^I+m^{I0}\, B_0)\wedge
(dA^I+m^{I0}\, B_0)-e_I{}^0\, B_0\wedge
(dA^I+\frac{1}{2}\,m^{I0}\, B_0)\,.\label{lg}\\
\end{eqnarray}
If one prefers to work in the second order formalism, it suffices
to use for $v^A$ the expression derived from eq. (\ref{gv}), all
the terms not containing $v$ in (\ref{lg}) being left unchanged.
The Lagrangian (\ref{lg}) is clearly invariant under the following
tensor--gauge transformation: \begin{eqnarray} \delta B_0&=&
d\xi_0\,\,\,;\,\,\,\delta B_A= D\xi_A\,\,\,;\,\,\,\,\delta
A^I=-m^{I0}\,d\xi_0\,\,\,;\,\,\,\,\delta v^A=0 \,,
\end{eqnarray}
and vector--gauge transformations:
\begin{eqnarray} \delta B_0&=&\delta B_A=\delta v^A=0\,\,\,;\,\,\,\,\delta
A^I=d\lambda^I\,.
\end{eqnarray}
 The introduction of the remaining
charges is more problematic and is left to future investigations.
\section{Concluding remarks}
In this paper we have investigated different aspects of Calabi--Yau compactifications to four dimensions in the presence of fluxes and their interpretation in terms of $N=2$ massive supergravities. When the quaternionic manifold is of special type, as it 
occurs in $N=2$ theories encompassing these compactifications, the scalar potential in the presence of cubic special geometries 
has a dual no--scale structure for certain choices of fluxes. These choices allow to reproduce on general grounds 
some computations of the scalar potential coming from half--flat manifolds or theories with NS flux \cite{gm,glmw,halfflat}.\par
In an attempt to introduce further fluxes, corresponding to magnetic charges, it is natural to consider a dual form 
of the special quaternionic geometry in which the Heisenberg scalars are replaced by antisymmetric tensors.
This results in a non--polynomial tensor theory \`a la Freedman and Townsend \cite{ft} which has a manifest symplectic invariance under the tensor--gauge transformation given in (\ref{db1st}) and (\ref{db2nd}). We note that the non--polynomial antisymmetric tensor theory
only contains as scalars the dilaton $\varphi$ in addition to NS Calabi--Yau $h_1$ complex deformations encoded in the special geometry of the original special quaternionic manifold. When this system is coupled to vector multiplets, a manifest ${\rm Sp}(2\, h_1+2)\times {\rm Sp}(2\, h_2+2)$ of the mirror special geometries is exhibited. The possibility of adding electric and magnetic fluxes in this framework will be discussed elsewhere. 
\section{Acknowledgments}
 R.D., M.T. and S.V.  would like to thank the Physics
Department of CERN, where part of this work was done, for its kind
hospitality. S.F. would like to thank the Physics Department of Politecnico di Torino for its kind hospitality and the restaurant 
``Le Tre Galline''.
\par
 Work
supported in part by the European Community's Human Potential
Program under contract MRTN-CT-2004-005104 `Constituents, fundamental forces and symmetries of the universe', in
which R. D'A. and M.T.  are associated to Torino University. The work of S.F.
has been supported in part by European Community's Human Potential
Program under contract MRTN-CT-2004-005104 `Constituents, fundamental forces and symmetries of the universe', in
association with INFN Frascati National Laboratories and by D.O.E.
grant DE-FG03-91ER40662, Task C. The work of S.V. has been
supported by DFG -- The German Science Foundation, DAAD -- the
German Academic Exchange Service and by the European Community's Human Potential
Program under contract  HPRN-CT-2000-00131.

\end{document}